# Generalized Functions & Experimental Methods of Obtaining Statistical Quantities Which Determine Preferences in Choice-Rich Environments

By *Leonid A. Shapiro*

"Need for mathematical basis should dominate one's search for any new theory. Philosophical or physical ideas must be adjusted to fit mathematics. Not other way around. One ought to start with mathematical basis. One can tinker with philosophical or physical ideas to adapt them to mathematics, but mathematics cannot by tinkered with. It is subject to rigid rules and is restricted by logic. Distrust philosophical or physical concepts as basis for theory. One should concentrate on getting interesting mathematics and put one's trust in it, even if it does not appear at first sight to be connected with physics. Equations of any theory are worked out before their physical meaning is obtained. Physical meaning follows behind mathematics. Physical interpretation is obtained only after mathematical basis is obtained" (Dirac 1978; abridged). Distributions such as $\delta x = 0$ for $x \neq 0$ where $\int_{-\infty}^{\infty} x \cdot \delta x = 1$, for instance, are not functions, but they are mathematical basis of physics (Dirac 1927).

**Abstract.** Preferences of individuals are distributions of elements generated by generalized functions. Models of economic decision-making derived from such distributions are consistent with results of physiological experiments, and explain any behavioral situations without simplifying assumptions. Quantities in such models precisely correspond to experimentally obtainable physiological observables which determine statistical properties of central nervous system as it represents different stimuli. Graphical method of consistently and quantitatively at-a-glance interpreting or visualizing physiological data within context of economic models is demonstrated. [77 words]

## §1

Sensation of person $Ч \in П$ classifies geometrical locations $L \in E^*$ of external physical environment $E^*$ into equivalence-classes $Э \in E$, creating external perceived environment $E$, whose geometry elements are variable-quantities $\xi \in \mathbb{R}$ defined on *intersections* of equivalence-classes or "things" $Э \in E$. "We cannot perceive or sense absolute change. We can only perceive or sense relative difference, or change from past configuration" (Boskovich 1755). So, if $JF = FJ = F$ and $JN = N$ (Menger 1944), then measurement $M : E \to \mathbb{R}$ assigns distinguishable "labels" or "stimuli" $\{\xi_J\}[N = \{-\infty,...,\infty\}] = \{\xi_N\}_{N=-\infty}^{\infty}$ as $N \to \infty$, which are different unique real-numbers, to different unique boundaries at which some equivalence-classes *transition* to other equivalence-classes, and thus, are distinguishable one from other. For instance, drop at table-edge is boundary *transition* by which floor is distinguished from table-top.

One-and-only criterion of mapping $M$ is distinguishability of real-numbers $\{\xi_J\}[N=\{-\infty,...,\infty\}] = \{\xi\} = \{..., \xi', \xi'', \xi''', ...\}$. They are different "labels" or "names." They are combined *without ambiguity* according to mapping $M$ with different intersections of equivalence-classes $\{Э_J\}[N=\{-\infty,...,\infty\}] = \{Э\} = \{..., Э', Э'', Э''', ...\}$, because such intersections are different boundaries of "things." Different "labels" or "names" correspond to *transition* from one quantity of "things" to different quantity of identical "things," *transition* from one quantity of "things" to identical quantity of different "things," and so on. Any sequence of changes in boundary *transition* context of environment, along or within future null cone, which depolarize or hyperpolarize nerve-cells, "stimulate" receptive-fields of nerve-cells, or "excite" nerve-cells, are different "things," e.g., different goods or money at different times. Ordered pair $(\xi_{J+\upsilon} - \xi_{J-\upsilon}, \xi_J - \xi_{J-\upsilon})$ is meaningless if $[\xi_{J+\upsilon} - \xi_J < 0] \cup [\xi_{J+\upsilon} - \xi_J > 0] \cup [\xi_J - \xi_{J-\upsilon} < 0] \cup [\xi_J - \xi_{J-\upsilon} > 0]$.

Different behavior or responses of person $Ч \in П$ to different stimuli $\{\xi\}$ correspond to differences between values $\{V_J\}[N=\{-\infty,...,\infty\}] = \{V\} = \{..., V', V'', V''', ...\}$, which are mapped $F: \mathbb{R} \to \mathbb{R}$ to different stimuli $\{\xi\}$, establishing ranking order. Differences in valuation are differences in attention or perception, and pleasure is "organizing" change or "growth" in state of brain neural network $M$ at synapses due to excitations arising from stimuli (Hebb 1949). For instance, growth in linkages within frontal areas of thalamus and connected areas of cerebral cortex would change intentional responses to search and consume problems, such as search for economic solutions, and simultaneously give pleasure (Pribram 1960). Different stimuli $\xi \in \mathbb{R}$ have different values $V \in \mathbb{R}$ mapped by $F$ onto them by different people $Ч \in П$, and so, different people exhibit different responses to identical stimuli or identical responses to different stimuli. Their behavior changes over time, because $F$ changes over time due excitation-initiated growth at synapses. "Preferences" are merely *different stimuli*; they are not "rational" at all; they cannot be chosen; neither helping other people at cost to self nor satisfying self exclusively is more or less "rational"; pleasure is change of state of brain by excitation-initiated growth of synapses caused by different sequences excitations, whose only physiological qualities are their distinguishability, novelty, and more or less organizing effect on already existing state of neural network, so that some order of stimuli is "preferred"; people cannot do something contrary to pleasure, as opposed to because of pleasure (Leibniz 1700; Hebb 1949).

Pain is evoked if stimuli are encountered in order that is different from preferences, i.e., "preferences are not satisfied." Preferences are not magnitudes, but topological changes in neural network of brain.

Environment which Ч senses is single geometrical manifold whose elements are $\xi$, in space $E$, and "wraps" around "things." Different qualities of distinguishable things are indicated by sets in neighborhood of each point $M^{-1}\xi \in E$, to which each point belongs. Physical environment $E^*$ underlying $E$ may be very different from what Ч perceives. Physical things may have geometries and qualities which are very different from what Ч thinks, but Ч behaves only according to differences between elements which their senses distinguish. Everything else has no influence on their behavior. They do not know about it. Things, insofar as they determine behavior of Ч, are merely different deformations of single geometrical manifold, which extends only insofar as senses and thoughts of Ч extend, and which envelopes Ч or surrounds them from all directions, at any moment. If small perturbation away from starting point in our single geometrical manifold leads to any other point which is in identical equivalence class as starting point, i.e., all points in neighborhood of starting point are in single equivalence-class, then neighborhood of starting point was "stable," so that small change in stimulus leads to small change in response. If several or infinitely many equivalence-classes intersect in neighborhood of starting point, then small perturbation from starting point in one set lead to another point which is in different equivalence-class, which is called "bifurcation," so that small change in stimulus leads to great change in response.

To anticipate or describe actions or behavior of set of people {..., Ч' $\in \Pi$, Ч" $\in \Pi$, Ч''' $\in \Pi$, ... }, we analyze sums of $F$'s or co-chains of commutative functions defined on $Э' \cap Э"$, $Э' \cap Э" \cap Э'''$, ..., $Э' \cap Э" \cap Э''' \cap ... \cap Э_P \in E$, if they are not empty, where $Э'$, $Э"$, $Э'''$, ..., $Э_P \in \{Э\}$, where $\{Э\}$ is set of coverings of $E^*$, and where $Э' = \{Э\}/[Э" \cup Э''' \cup ... \cup Э_P]$, $Э" = \{Э\}/[Э' \cup Э''' \cup ... \cup Э_P]$, $Э''' = \{Э\}/[Э' \cup Э" \cup ... \cup Э_P]$, ..., $Э_P = \{Э\}/[Э' \cup Э" \cup ... \cup Э_{P-\upsilon}]$, up to $P$-fold intersection, because class of ordered pairs $(M, F)$, which defines class of ordered 3-tuples $(Э, \xi, V)$, determines actions or behavior (e.g., Eilenberg & Steenrod 1952; Wells 1980). Then, we ask: What "action" or tangent-space $Б \in E$ or what "behavior" or tangent-space $Б^* \in E^*$ maximizes $F\xi \in \mathbb{R}$, given "controls" or preferences of person Ч $\in \Pi$? (e.g., Rosen 1980)

Let $J\xi = \xi$ and $\mathcal{A}_P(\xi_A,...,\xi_B) = \xi_P$, so that labels $\mathcal{A} = \{\mathcal{A}_A, ..., \mathcal{A}_B\}$, where measurement $M : E \cap \mathrm{o} \to \mathbb{R}$ maps different intersections of $\{\Im\}$ *which are removed*, according to rules $A, ..., B$, to real numbers $\xi_A, ..., \xi_B$, respectively. In environment $E$, all changes have corresponding variable-quantity labels in this way. Let $v = \{..., v', v'', v''', ...\} \cap [\mathbb{R} \cup \mathbb{C}]$. If $\mathcal{D} = d/dJ$, $\mathcal{D}^{v_P} = d^{v_P}/dJ^{v_P}$, $\mathcal{D}_{\mathcal{A}_P} = \partial/\partial\mathcal{A}_P$, $\mathcal{D}_{\mathcal{A}_P}^{v_P} = \partial^{v_P}/\partial\mathcal{A}_P^{v_P}$, and operator $O = \sum_{\mathcal{A} \cup v} \mathcal{K}_{\mathcal{A} \cup v} \cdot J$, then $O\mathcal{D}^v = \sum_v \mathcal{K}_v \cdot \mathcal{D}^v$ and $O\mathcal{D} = \sum_{\mathcal{A},v} \mathcal{K}_{\mathcal{A},v} \cdot \mathcal{D}_{\mathcal{A}}^v$. "Information" about values, in sense that we require it, is "influence structure" of $F$, which is set of answers of kind $\{\mathcal{D}_{\mathcal{A}} F\}$ to following set of questions: if $\mathcal{A}$ changes, what happens to $F$? (Rosen 1984)

If several simultaneous changes in environment are investigated, then we seek set of answers of kind $\{\mathcal{D}_{\mathcal{A}_A}^{v'} ... \mathcal{D}_{\mathcal{A}_B}^{v'''} F\}$. In fact, any question we are interested, however complicated, in may answered by selecting coefficients $\{\mathcal{K}_j\}[N = \{-\infty,...,\infty\}] = \{\mathcal{K}\} = \{..., \mathcal{K}', \mathcal{K}'', \mathcal{K}''', ...\}$, labeling environmental changes, and computing: $\sum_{\mathcal{A}} \sum_{v=-\infty}^{\infty} \mathcal{K}_v \cdot \mathcal{D}_{\mathcal{A}}^v F$, which is arbitrary part of "influence structure" of $F$. And so, we need to know how to determine $F$ in most general way.

## §2

Mapping $F : [\xi \in \mathbb{R}] \to [V \in \mathbb{R}]$, which is continuous or dispersed, of stimuli $\{\xi\}$ to values $\{V\}$, is generated as right-hand limiting sum across real boundary, by equivalence-class $F$ of pairs of analytic functions, $[F_+ : [\xi \in \mathbb{R} + i\cdot\sigma \in \mathbb{C}] \to V_+ \in \mathbb{R}, F_- : [\xi \in \mathbb{R} - i\cdot\sigma \in \mathbb{C}] \to V_- \in \mathbb{R}]$, which are different flows in complex domains above and below real boundary, respectively, where $F(\xi \in \mathbb{R}) = [F_+(\xi + i\cdot\sigma), F_-(\xi - i\cdot\sigma)] = \lim_{\sigma \to 0+} F_+(\xi + i\cdot\sigma) - F_-(\xi - i\cdot\sigma) = [V \in \mathbb{R}]$, and because many different pairs of analytic functions generate identical distribution of values in such way, we may always choose $[F_+, F_-]$ so that hierarchical order of values $\{V', V'', V''', ...\}$ is not changed by transformation $\varepsilon F \rho$ (Kothe 1952; Sato 1958; Schapira 1972; Isao 1992).

Let re = $[\mathbb{R} \cap J]$, arg = $[\mathbb{C} \cap J]$, and $\mathcal{P}(D(S))$ be ring of all regular analytic functions in open complex-neighborhood $D(S) \in \mathbb{C}$ of open section $S \in \mathbb{R}$ restricted to real-line. Let, then, $\mathcal{P}(D(S)/S)$ be ring of all regular analytic functions in complex neighborhood of $S$ excluding those exactly on $S$. If $\mathbb{C}_+ = \{\kappa \in \mathbb{C} \cap \arg \kappa > 0\}$, $\mathbb{C}_- = \{\kappa \in \mathbb{C} \cap \arg \kappa < 0\}$, then equivalence-relation $[F = F_+$ if $D \cap \mathbb{C}_+$, $F_-$ if $D \cap \mathbb{C}_-] \in \mathcal{P}(D(S)/S)$ is generalized function or hyper-function, and $\mathcal{B}(S) = \mathcal{P}(D(S)/S)/\mathcal{P}(D(S))$ is set of every hyper-function defined on $S$ (Kothe 1952; Sato 1958; Sato 1959). A sheaf co-homology definition of $\mathcal{B}(\mathcal{M})$ is required for set of every hyper-function defined on real-manifold $\mathcal{M}$, or sum of such manifolds (which is required for analysis of behavior of individual person derived from function $F$, and for behavior of many people), but this is treated in §3 of this paper, whereas §2 of this paper considers function $F$ for one person (Sato 1960; Kashiwara, Kawai, & Sato 1973).

Equivalence-class nature of generalized functions, encompassing many pairs of components, allows us to require that hierarchical order of values is not changed by conformal re-scaling $\widehat{F} = \Omega \cdot F$ of metric corresponding to conformal re-scaling $\{\widehat{V}_J\}[N = \{-\infty,...,\infty\}] = \{\Omega_J \cdot V_J\}[N = \{-\infty,...,\infty\}]$ of values, in space $E$. Thus, ordered pair $(V_{J-\upsilon} - V_J, V_J - V_{J+\upsilon})$ is meaningless if $(V_{J-\upsilon} < V_J < V_{J+\upsilon}) = (\widehat{V}_{J-\upsilon} < \widehat{V}_J < \widehat{V}_{J+\upsilon})$. Magnitudes may be increased or decreased, in any way, if *angles* between geometrical elements, which represent values in metric space, are preserved (e.g., Penrose 1964; 1968). That is, $F\xi$ is marginal utility distribution or anticipated preferences. Of course, preferences are always merely expected, because action now can only satisfy wants in near or distant future when desired effect appears: real hierarchical ranking of distinguishable stimuli is limiting sum of two continuous functions of complex variable-quantities, $\xi + i \cdot \sigma$ and $\xi - i \cdot \sigma$, respectively, and if $\xi + i \cdot \sigma$ and $\xi - i \cdot \sigma$ have physiological meaning, then activation-inhibition network of nerve-cells may be placed in quantitative correspondence with choices people make. These results all follow trivially from well-known theorems about generalized functions, hyper-functions, which generate distributions, and micro-functions, which describe properties of singular points of such distributions:

1. every dispersed set generated by mapping $F$, including any set which is not any kind of function, say, because its lacks *consistent* segmented "motion" $(\zeta(\tau), \eta(\tau)) \in F$ transitioning from every $a \in A$ to $b \in B$ in $F: A \to B$, $a, b \in \mathbb{R}, \mathbb{C}$ (Menger 1954), e.g., hierarchical ordering $V' < V''$ in dispersed set-of-values $\{V\}$, where metric $F$ determines distance between $V' \in \{V\}$ and $V'' \in \{V\}$, is invariant to change in distance $V' - V''$ arising from of expansion $\varepsilon F$ or from restriction $F\rho$ (Cuhel 1907; Bernardelli 1934; 1936; 1938; 1939; 1952), may be treated as *distribution* of elements in real-space without causal quantitative connection in real-space, but with underlying quantitative causal connection in complex-space;

    2. every distribution of single real input may be generated as limiting sum of two analytic functions, and differentiation or integration may be applied to such functions; Leibniz's Law of Continuity is vindicated mathematically, no meta-mathematical assumptions being necessary;

    3. every two different generating functions of most general kind, $F'$ and $F''$, have addition defined for them, but multiplication of two different generating functions is not defined at all, and neither $F'\cdot F''$ nor $\log F' + \log F'' = \log(F'\cdot F'')$ is defined, so that phase-space with common degrees of freedom does not exist in most general cases, prohibiting inter-personal comparisons of values for lack of common phase-space, no extra-mathematical assumptions being necessary (Sato 1958; Kashiwara, Kawai, & Sato 1973; Kashiwara 1980; Isao 1992).

*What precisely are complex-number $\kappa = \xi + i\cdot\sigma$ and complex-conjugate $\bar{\kappa} = \xi - i\cdot\sigma$ in human central nervous system?* Instead of trying to physically interpret $\kappa$ and $\bar{\kappa}$ directly, let us "encode" $\kappa$ and $\bar{\kappa}$ onto physiological-"indicators" $\alpha$ and $\beta$, respectively, so that if $\varphi\uparrow$ $\uparrow\phi$, then

$$\begin{array}{ccc} \alpha \ldots + \ldots \beta & E^* \xrightarrow{M} V(\alpha,\beta) \xrightarrow{F(\varphi^{-1},\phi^{-1})} (V,E) \\ \updownarrow & (\varphi,\phi)\uparrow \quad\quad\quad \updownarrow \\ \kappa \ldots + \ldots \bar{\kappa} & E^* \xrightarrow{M} V(\kappa,\bar{\kappa}) \xrightarrow{F} (V,E) \end{array}$$

commutes (e.g., Rosen 1962; 1963; 1983). If we may physically interpret $\alpha$, $\beta$, $\varphi^{-1}$, and $\phi^{-1}$, then our analysis may proceed.

    At first, we may try something else: for indices $I \in \mathbb{R}$ and ordered $N$-tuples or "bundles" $(Q_{1,I},...,Q_{N,I} \in \mathbb{R})$ of various quantities of different goods $\{G_1,...,G_N \in E\}$ satisfying $U(Q_{1,I},...,Q_{N,I},I) = 0$ for indifference-curves $\Phi(Q_{1,I},...,Q_{N,I}) = I$, if $(dQ_{1,I}/d\tau, ..., dQ_{N,I}/d\tau) = (0,...,0)$, and $\nabla\Phi(Q_{1,I},...,Q_{N,I} \in \mathbb{R}) > (0,...,0)$, then person $Ч \in \Pi$ values "bundles" corresponding to $I'$ less than "bundles" corresponding to $I''$ if and only if $I' < I''$ (Pareto 1906a; 1909). As best explained by original inventor of this method, surprising problems *prevent* this method from describing most behavioral situations in our world.

"If order of consumption of several economic goods in any bundle is matter of indifference, one has system of indices of marginal utility which depend on quantities consumed. If order of consumption is not matter of indifference, one has system of indices of marginal utility which depend on consumption-paths followed. One individual, following some path during consumption of several economic goods in any bundle, starts at point $(Q_{1,I},...,Q_{N,I})$, and moving from that point, consumes quantities $(dQ_{1,I},...,dQ_{N,I})$. We assume: any individual chooses that consumption-path which gives this individual maximum satisfaction: first they consume $Y$, second they consume $X$, and so on. We shall assume that marginal utility does not dependent upon order of consumption. When marginal utility is not-dependent on order of consumption of goods, and is dependent only quantities of goods that are consumed, we can determine marginal utility" (Pareto 1906b; abridged). "If pleasure arising from consumption of $dX$ only depends on $X$, pleasure arising from consumption of $dY$ only depends on $Y$, and so on, or if pleasure arising from consumption of $(dQ_{1,I}, ..., dQ_{N,I})$ is different according to order of consumption of $dQ_{1,I}, ..., dQ_{N,I}$ but we admit that we may experimentally determine and fix order of consumption of $dQ_{1,I}, ..., dQ_{N,I}$, then we can construct indifference-functions as indices of choices. If order of consumption influences choices, we must necessarily fix order of consumption of $dQ_{1,I}, ..., dQ_{N,I}$ before we can determine points of equilibrium or function of $(Q_{1,I},...,Q_{N,I})$ which serves as index of choices" (Pareto 1909; abridged).

What if bundles are expanded to include every distinguishable equivalence-class $\ni \in E$, or situations represented by stimulus $\bigcup_{\lambda=1}^{\infty} \xi_\lambda \in \mathbb{R}$, as opposed to any restricted stimulus? Our world, without simplifying assumptions, is such a "choice-rich" environment, which introduces differences between predictions of any such economic model and behavior predicted.

We cannot assume that order of consumption of perhaps infinitely many different quantities of different goods, within single bundle of goods, does not affect pleasure achieved from consuming such bundle of goods. In choice-rich environment, bundles of goods are supposed to be hierarchically ranked, and yet value of bundles is not defined unless order of consumption is defined, but order of consumption, is not fixed ahead of time, physiologically. Order of consumption is simultaneously determined with bundles, and change in it determines change in value of bundles equally as much as change in quantities under consideration, but analysis of bundles of quantities leaves out analysis of order of consumption. If both are considered, then bundles no longer exclusively determine values, and we get dispersed set of hierarchically ordered real-numbers, which if it also supposed to be invariant under transformations, is not tractable in real-space if we wish to do further analysis. New mathematical basis is required. If we eat our food standing up, and then sit down, or if we sit down first and then eat our food, are these not different stimuli which give us different pleasure, and to which we react differently? If we must sometimes do one, and sometimes do other, our model must describe both situations differently. Only if we answer such a question for choice-rich environments, is any model capable of giving behavioral predictions in physical environment $E^*$ in which we really live.

Human central nervous system is activation-inhibition network of nerve-cells where excitations evoke changes at synapses according to processes at level of dendrites which are superposed over inhibitory background of changing magnitude (Rashevsky 1933; 1934; Hebb 1949; Hayek 1952; Pribram 1986; 2003). Activity and excitation, of course, increases temperature of activation-inhibition network (Gerard & Serota 1938; Serota 1939). Also, volume diffusion transmission of free molecules causes bulk activation or inhibition, or bulk change in rate of activation or inhibition, at lower energy cost, wherever memory-related precision of activation-inhibition networks, is not required, e.g., readiness for motion, sensation of hunger, and so on (Bach-y-Rita 1993; 2005). Individual nerve-cells are spontaneously excited over time with regular frequency generated underlying wave-like processes in dendrites, or may be above and beyond excited or inhibited by environmental stimuli or connected nerve-cells (Rashevsky 1933; 1934; Gerard & Young 1937; Gerard & Libet 1939; 1941).

There is significant delay between when stimulus is and begins to cause modification of neural network and response and when it consciously felt, sensation being evoked only as dendritic changes as synapses occur, and vice versa, motivation is always unconscious, whereas consciousness, if present, is only ability to interrupt motivation, given later final state of neural network; earlier stimuli may be masked or enhanced by later stimuli according to clock time, and due to unconscious processing, internal time has no 1-to-1 correspondence with clock or proper time whatsoever; most stimuli are not consciously felt at all, but elicit unconscious conditioned or orienting responses (Condillac 1754a; 1754b; Sokolov 1958; Deecke & Kornhuber 1965;

Libet 1966; 1973; 1981; 2003; Alberts et al 1967; Feinstein et al 1979; Brass, Haynes, Heinze, & Soon 2008).

Central nervous system covers geometry of physical environment $E*$ with distinguishable equivalence-classes $Э \in E$. It may only code, experiments show, however, auditory, tactile, and visual excitations in response to stimuli distinguished by neuronal model $M$, as probability-distribution functions, which are fully described predicted by postulating a barrier beginning at certain height which fluctuates while being approached by random-walk motion as function of proper time $\tau$ parameterized by a positive drift-coefficient; at that point, one nerve-cell emits signal to any other connected nerve-cell, so that individual time-intervals between excitations are stochastic, code no information, and instead, stimuli distinguishable by neuronal model are represented as distinguishable statistical information about time-intervals, namely, probability-distribution functions with different combinations of barrier and drift-coefficient which generate such time-intervals: "neuron is thus processor of stochastic dendritic events which displays its computed output as statistics of sequence of inter-spike intervals" (Berger et al 1990; Berger & Pribram 1992; 1993; King, Pribram, & Xie 1994).

"Attention" or selective perception of simultaneously present stimuli of different kinds, and "anticipation," or speculative responsiveness to expected stimuli has been know to exist, but anatomical apparatus involves in such processes occurred was not know (Condillac 1754a; 1754b). Actually, in mammals: afferent nervous fibers in optic nerve transmit excitations from retina to visual cortex in the brain, but efferent fibers optic nerve transmit excitations from central nervous system to retina, changing excitation pattern response to environmental stimuli in the retina as function of neuronal model $M$; for instance, visual stimuli which have already ceased to exist, but earlier were significant cause of excitations, evoke responses in the retina, and auditory and tactile stimuli, which modify neuronal model, then modify excitation patterns in the retina whose cause is visual stimuli (Brouwer 1933; Pribram, Spinelli, & Weingarten 1965; Spinelli & Weingarten 1966a; 1966b). Such "centrifugal control" of excitation of nerve-cells is present in muscle-cells and many periphery nerve-cells or receptors are excited or inhibited by environment, which have both afferent fibers leading away from them to brain stem and central nervous system and efferent fibers leading to them from brain stem and central nervous system, permitting self-stimulation at periphery, eliminating possibility of any "pure" experience controlling learning, and allowing environmental noise to be filtered out by central nervous system control of firing of muscle-cells and periphery nerve-cells (Granit 1955). Stimuli which are present may be ignored, evoking no excitations beyond periphery nerve-cells. Stimuli which are absent may be perceived, evoking many excitations beyond periphery nerve-cells. As neuronal model $M$ changes, which stimuli evoke which sensations changes; whether or not two different stimuli evoke sensation of single stimuli or two distinguishable stimuli depends on state of neuronal model $M$ (Hayek 1952; Sokolov 1960a; 1960b). We see, then, that:

1. *$\alpha$ must be height of barrier that must be reached by random-walk model before nerve-cells pass excitation threshold and fire, while $\beta$ must be drift-coefficient in random walk model*; these experimentally measurable variable-quantities and only these are found to fully describe *representation of stimuli in any way as responses of $M$ to stimuli, taking every other part of $M$ already into account*; there are no other alternatives which have not been excluded;

2. $\varphi$, $\phi$, $\varphi^{-1}$, $\phi^{-1} \in M$; in humans, $\varphi$ and $\phi$ are probably determined by activation-inhibition network in temporal cortex, which categorizes stochastic processes underlying imaging (Pribram 1991), but $\varphi^{-1}$ and $\phi^{-1}$ are likely determined by activation-inhibition network in frontal cortex, because lesions there much significantly than lesions in posterior cortex impair ability to notice and use visual cues or to associate emotions, such as humor, with sensations of stimuli in order to then use those associations to recall sensations (Pribram & Prigatano 1981);

3. network of linkages of nerve-cells has statistical metric (Menger 1942), because of statistical nature of information and perception as it exists within central nervous system of humans and other mammals.

So, then, "pleasure is not activity of particular structures in nervous system, nor is it particular kind of pattern of cerebral organization; pleasure is growth or development of cerebral organization" (Hebb 1949). "Satiety" or "emotion," which determines when behavior "stops" is mostly determined in forebrain, in amygdala; "motivation," which determines when behavior "begins," when to "go," or when to "continue going" is mostly determined in striatum of basal ganglia (Hebb 1949; Pribram 2003). Cerebral cortex determines actions, because it determines memory (and thus, perception, pleasure, pain), but it neither initiates behavior, which is done in basal ganglia hippocampus and striatum, nor interrupts behavior and so that other actions may begin, which is done in amygdala, and then, hippocampus determines "attitudes" or classes of ordered pairs of motivation and emotion (Pribram 2006).

We see that rankings of equivalence classes which are mapped by central nervous system to its perceived environment are topological, i.e., there is quantitative measure of "nearness" or "distance" between elements in space. Sensation of stimulus diminishes if stimulus is repeated, because "memory" or neuronal model $M$ of anticipated stimuli, environment, and self is constructed over time by experience through connections and positions of nerve-cells in cerebral cortex, and maps particular sensations to particular stimuli, what parts of environment are perceived and how they are perceived, but widespread lesions to cerebral cortex do not destroy any part of neuronal model already formed; "memory" does not change after lesions, but lesion significantly decrease rate of "learning," i.e., rate at which neuronal model $M$ changes (Lashley 1929; Sokolov 1958; 1960a; 1960b; Pribram 1966). When lesions occur, functions of destroyed parts of cerebral cortex, periphery nerves, and visual cortex, unless quantitative damage is very extensive, are moved to another part of brain and gradually return, e.g., "we see with brain, not with eyes" (Lashley 1929; 1948; Frank & Lashley 1934; Hayek 1952; Bach-y-Rita 1967; Bach-y-Rita & Collins 1970; Bach-y-Rita, Tyler, & Kaczmarek 2003).

Thus, "nearness" and "distance" quantitatively determine structure of central nervous system, structure of neuronal model, and products of metabolic and repair processes occurring in central nervous system, but pleasure or pain evoked by any stimulus are totally invariant to quantitative structure and changes in quantitative structure; gross destruction and change in quantitative structure experimentally produce no change in memory, so that pleasure or pain evoked by any environmental stimulus is determined only by change or growth of connections which determine relative position of elements in space, but not upon quantitative differences in relative position of elements in space (Lashley 1929; Hebb 1949; Hayek 1952). Metric $F$ is required to indicate "nearness" and "distance" between equivalences-classes of distinguishable things in perceived environment (Rosen 1976; 1977) while satisfying above constraints. This results in hierarchical and ordinal nature of preferences concerning all different permutations of distinguishable stimuli (Cuhel 1907). Cost, then, is "regret of forgone opportunity," namely, next greatest preference in immediate future which could have alternatively been satisfied by behavior, and must be forgone in immediate future if greatest preference, and present behavior is carried out instead of it (Condillac 1754a; 1754b).

Supposedly, activation-inhibition networks sometimes generate "heterarchies" of values, $F(\xi') > F(\xi'') > F(\xi''') > F(\xi')$, where some entity prefers state $\xi'$ of perceived environment

*E* instead of $\xi''$, prefers $\xi''$ instead $\xi'''$, but prefers $\xi'''$ instead of $\xi'$, e.g., mouse prefers food over sex, sex over avoidance of electric shock, and avoidance of electric shock over food (McCulluch 1945a; 1945b; 1956). Actually, "coarse" measurement which records physically and perceptually distinguishable changes in quantities of things or different stimuli which evoke distinguishable responses as if they were identical changes in quantities of things or identical stimuli, makes "heterarchies" of values appear to exist. Any "coarse" topology $F(\xi') > F(\xi'') > F(\xi''') > F(\xi')$, and hypothetical activation-inhibition network which generates it, through "refinement" of measurement, is revealed to be "finer" topology $F(\Omega'\cdot\xi') > F(\Omega'\cdot\xi'') > F(\Omega'\cdot\xi''') > F(\Omega''\cdot\xi' = \Omega'\cdot\xi''')$ with corresponding actual activation-inhibition network (e.g., Rosen 1976; 1977). And this is why David Hume warned that any subjects of inquiry must be quantified: bread is never preferred to water, and water is never preferred to bread; concrete changes in quantities of bread are preferred to concrete changes in quantities of water, but other changes in quantities of water are preferred to other changes in quantities of bread, which is of course old news (North 1690; Condillac 1776; Lloyd 1834; Gossen 1854; Jevons 1871; Menger 1871). So, pleasure is "network of neurons is becoming more and more organized," pain is "disruption" of such increasing organization, and emotion is "disruption" of "behavior," either to prolong or preserve "pleasant stimulus" or to avoid or remove "painful stimulus" (Hebb 1949).

For instance, consider one alternative model of preferences and how it is contradicted by such considerations. Possible actions $A, B, C, D$ are chosen over time, so that $Ч \in П$ chooses *A* with probability $P(A,B)$ and *B* with probability $P(B,A) = 1 - P(A,B)$, so that if $P(A,B) > 1/2$, then *A* is preferred to *B*, leading to utility-function *U* according to which $[P(A,B) > P(C,D)] = [U(A) - U(B) > U(C) - U(D)]$ (Debreu 1958). Probability is frequency, or infinite wave-train, or, at least, precisely identical choices must be repeated many times for function *P* to be defined (Gabor 1946). What contradicts any possibility of function *P* and corresponding function *U*, in such sense, is that activation-inhibition network is being modified at synapses at level of dendrites after every action and sensation, and inhibitory background, treated as field, is also changing after each and every action and sensation; every situation where choices must be made is unique, and pleasure is "growth" of neural network, which is precisely how learning occurs (Hebb 1949; Pribram 1991). Principle of "computational equivalence" means, furthermore, that we *cannot* simulate *a priori* repetition of events in *E\** and *E*, which actually happen uniquely insofar as $M$ is concerned, to obtain $P$ and $U$, because complex systems cannot have their future output computed in shorter way than trajectory in their underlying dynamics required to obtain such output; "no shortcut is possible" (Hayek 1952; Rosen 1962; Wolfram 1985; Wolfram 2002). "Complexity" of system $S$ is measured by number $N$ of alternative models of system $S$ which are not isomorphic, all being required to explain behavior of system $S$ or all being different ways in which we may interact with system $S$, so that any complex system $S$ has bifurcations in behavior, e.g., people make radical changes in their trajectories (Rosen 1979). We measure complexity by $\log_2 N$, so that "simple" systems have "complexity" $\log_2 1 = 0$, while "complex" systems have "complexity" $\log_2 N > 0$. Metric $F$ lends itself to describing complex systems, which minds are, because it may generate sets not consistently related in real-space but only consistently related in complex-space.

By seeing how people actually rank different stimuli in terms of ordered pairs $(\alpha, \beta)$, we gain data about form of $F_+\varphi^{-1}$ and $F_-\phi^{-1}$. Whatever data we collect has $F_+\varphi^{-1}$ and $F_-\phi^{-1}$ fit to describe such data, because every distribution may be described in such way. Whole hierarchies of values of infinitely many different stimuli are distinguished and represented at-a-glance by assigning different diagrams with two degrees of freedom to different ordered pairs $(\alpha, \beta)$.

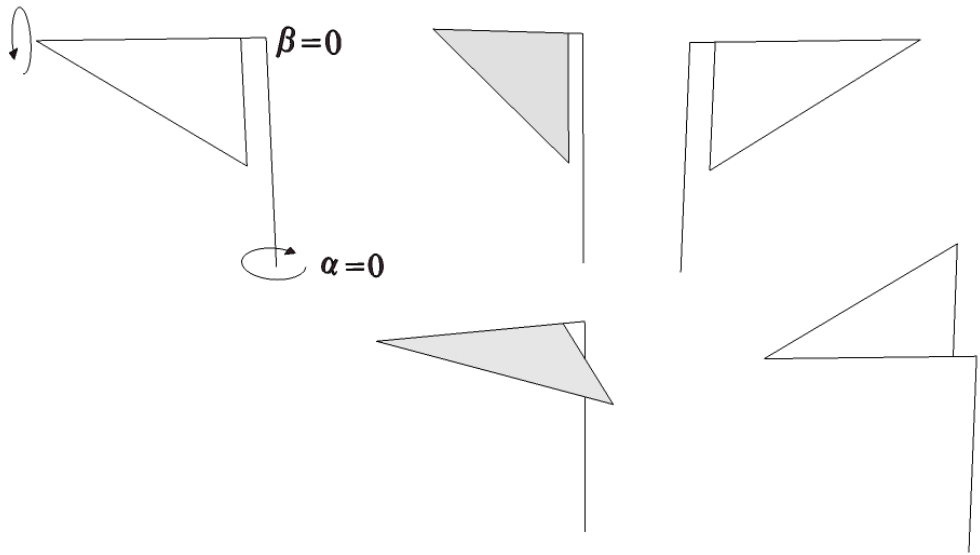

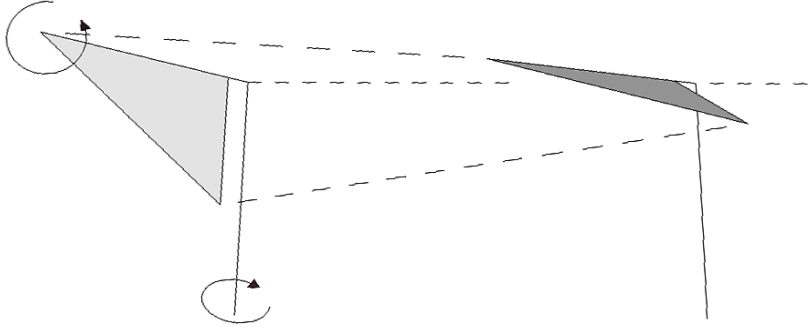

Figure 1. For any person, at-a-glance representations of different hierarchies of values. If any one ranking changes, every other ranking changes, because ranking is relative, which agrees with observed difficulty and discomfort of making decisions concerning stimuli encountered for first time (e.g., Wheeler 1920).

By associating different partitions $\bigcup_{\lambda=-\infty}^{\eta_1} \xi_\lambda$, $\bigcup_{\lambda=\eta_1}^{\eta_2} \xi_\lambda$, $\bigcup_{\lambda=\eta_2}^{\eta_3} \xi_\lambda$, ..., $\bigcup_{\lambda=\eta_3}^{\infty} \xi_\lambda$ to different diagrams of such kind, and thus seeing how observable-pairs which generate such diagrams are hierarchically-ranked or what order such ranges of diagrams, situations mapped to each of such classes, may be arranged in, i.e., by experimentally observing what order situations are preferred in, and knowing that such ranking are consistent for individual people, we learn something about $F_+\varphi^{-1}$ and $F_-\phi^{-1}$ which generate different rankings of stimuli for individual people when different classes of stimuli are present. We see how individual people rank or prefer different stimuli. By knowing that orienting reflex decreases when stimulus is repeated, due to growth at synapses resulting in lower resistance (one reason for decreasing marginal utility rule), we narrow down classifications. If $\{..., V', V'', V''', ...\}$ is set of preferences, where $... < V' < V'' < V''' < ...$ is hierarchy of values, and $\mathcal{X}\xi$ is equivalence-class of restriction homomorphisms of kind $R_{\rho''}^{\rho'}: V\rho' \to V\rho''$, where $\rho''$ is contained in $\rho'$, then values are outputs of kind of *step-function*, which is not continuous, with singular points in between ranges of inputs:

$$F = \begin{cases} ........, & .............., \\ V' \cdot \mathcal{X}J, & Я' < J < Я'' \\ V'' \cdot \mathcal{X}J, & Я'' < J < Я''' \\ V''' \cdot \mathcal{X}J, & Я''' < J < Я'''' \\ ........; & ................ \end{cases}.$$

It is possible to de-compose $F\xi$ into sum of hyper-functions, which is defined by contour integration (from $-\infty$ to $\infty$) of corresponding sum of impulses (Sato 1959; Isao 1992). *Infinite sum* (for all labels) of *projections* of *values*, which converges in complex-space, generates value hierarchy in real-space:

$$F\xi = \ldots + V'\cdot \mathcal{X}\xi \cdot \begin{cases} 1, & Я' < \xi < Я'' \\ 0, & \xi < Я', Я'' < \xi \end{cases} + V''\cdot \mathcal{X}\xi \cdot \begin{cases} 1, & Я'' < \xi < Я''' \\ 0, & \xi < Я'', Я''' < \xi \end{cases}$$

$$+ V'''\cdot \mathcal{X}\xi \cdot \begin{cases} 1, & Я''' < \xi < Я'''' \\ 0, & \xi < Я''', Я'''' < \xi \end{cases} + \ldots$$

$$= \ldots + V'\cdot \mathcal{X}\xi \cdot \frac{1}{j} \cdot \log_e\left(\frac{\kappa - Я''}{\kappa - Я'}\right) + V''\cdot \mathcal{X}\xi \cdot \frac{1}{j} \cdot \log_e\left(\frac{\kappa - Я'''}{\kappa - Я''}\right)$$

$$+ V'''\cdot \mathcal{X}\xi \cdot \frac{1}{j} \cdot \log_e\left(\frac{\kappa - Я''''}{\kappa - Я'''}\right) + \ldots$$

$$= \sum\nolimits_{Я_\lambda, Я_{\lambda+\varepsilon}} \left[\frac{V_\lambda \cdot \mathcal{X}\xi}{j} \cdot \log_e\left(\frac{\kappa - Я_{\lambda+\varepsilon}}{\kappa - Я_\lambda}\right)\right] \cap -\infty < Я_\lambda < Я_{\lambda+\varepsilon} < \infty$$

$$\cap \ i = \sqrt{-1} \ \cap \ j = 2\cdot\pi\cdot i$$

$$\cap \ (V_{J-\upsilon}\rho' < V_J\rho' < V_{J+\upsilon}\rho') = (V_{J-\upsilon}\rho'' < V_J\rho'' < V_{J+\upsilon}\rho'').$$

That is, because $\mathbb{C}_+ = \{\kappa \in \mathbb{C} \cap \arg\kappa > 0\}$ and $\mathbb{C}_- = \{\kappa \in \mathbb{C} \cap \arg\kappa < 0\}$, then

$$V = F\xi = \lim_{\sigma \to 0+} F_+\kappa - F_-\overline{\kappa} = \lim_{\sigma \to 0+} F_+\varphi^{-1}\alpha - F_-\phi^{-1}\beta =$$

$$\lim_{\sigma \to 0+}\left[\sum\nolimits_{Я_\lambda, Я_{\lambda+\varepsilon}}\left[\frac{V_\lambda \cdot \mathcal{X}\xi}{j}\cdot\log_e\left(\frac{\varphi^{-1}\alpha - Я_{\lambda+\varepsilon}}{\varphi^{-1}\alpha - Я_\lambda}\right)\right]\cap \mathbb{C}_+\right] - \left[\sum\nolimits_{Я_\lambda, Я_{\lambda+\varepsilon}}\left[\frac{V_\lambda \cdot \mathcal{X}\xi}{j}\cdot\log_e\left(\frac{\phi^{-1}\beta - Я_{\lambda+\varepsilon}}{\phi^{-1}\beta - Я_\lambda}\right)\right]\cap \mathbb{C}_-\right].$$

Then, two analytic functions exist, which are continuous and have derivatives defined for them, and we take $\mathcal{D}F$ (for any one restriction), so that solution $\xi^*$ is "decision" of what stimulus to respond to. That is, other stimuli are ignored until $\xi^*$ has been responded to, and such resulting behavior changes environment, presenting other stimuli, and next decision about what to respond to is then made in same way.

Singularities are locations in vector fields where directions of extensive magnitudes are not defined. Set of singular points $\{..., Я', Я'', Я''', ...\}$ are unique stimuli for which $Ч \in П$ has neither unique motivation nor unique behavior. It is set of unique stimuli for which $Ч \in П$ lacks unique responses. If such stimuli are encountered then instead of starting new motivation and its corresponding behavior in addition to present motivation and its corresponding behavior, emotion $\vartheta = J + ҊЯ \cap Я = J$ would be evoked, which is excitation and its corresponding movement that *disrupts* present behavior by adding $ҊЯ \in \mathbb{R}$ to $\xi \in \mathbb{R}$, depending which kind of emotion is evoked, so that

$$\xi(\tau+\upsilon) = \vartheta\xi\tau = \xi\tau + ҊЯ \cdot \begin{cases} 1, & Я = \xi\tau \\ 0, & \xi < Я, Я < \xi\tau \end{cases}.$$

Emotion causes $Ч \in П$ to stop present motivation and its corresponding behavior, merely allowing next or previous motivation and its corresponding behavior to be evoked (e.g., Hebb 1949, Pribram 2003; 2006). In not-conscious way, emotions-function $\vartheta$ "joins" together many separate parts or "patches" of memory into one preferences-function $F$, which determines conscious decision-making, and would explain why lesions in brain that impair association of emotions and sensations makes remembering sensations and re-constructing past or present sensations in future much more difficult (e.g., Hebb 1949; Pribram & Prigatano 1981). Emotion evolved, because very quick transition between some one behavior and some other behavior made survival of individual much easier and more likely in many situations. Emotion is necessary for long-term memory and complicated preferences. If, however, people have necessary information to make valid decisions, but make "mistakes" anyway, or if they "panic," this also is anticipated by analysis of singular points of $F$.

For identical stimuli (i.e., decoded labels $\xi \in \mathbb{R}$), if $Ҋ$ is index transformation or restriction which does not preserve hierarchical order $... < V' < V'' < V''' < ...$, then exchange or "trade" of things occurs between person $Ч \in П$ and person $ч \in П$ when $F_Ч = F_ч Ҋ$, i.e., when $F_Ч$ has all or some $V_\lambda$'s in reverse order of $V_\lambda$'s of $F_ч$, and extent of exchange is proportional to how many $V_\lambda$'s are associated with different impulses, or number of different contradictions in $F_Ч \cap F_ч$.

For person $Ч \in П$, if $[F(R_A < \xi^* < R_{A'}) = V_A] = [F(R_B < \xi^* < R_{B'}) = V_B]$, then what behavior is evoked? If system has unique solutions range, then it resists environmental gradient in any direction away from $\{\xi_J^*\}$ but simultaneously moves in direction of $\{\xi_J^*\}$. If system has several solutions ranges, then it resists environmental gradient in any direction away from $\{\xi_J^*\}$ but neither moves in direction corresponding to $F^{-1}V_A$ nor moves in direction corresponding to $F^{-1}V_B$, because choice cannot be made, for lack of reason, and but because reason exists for making one or other choice, however, it not resists environmental gradient in direction corresponding to $F^{-1}V_A$ or environmental gradient in direction corresponding to $F^{-1}V_B$, so that first perturbation of environment in direction of $F^{-1}V_A$ or in direction of $F^{-1}V_B$ determines choice.

Neuronal model $M$ exists in form of probability-distribution functions described by statistical metric (Menger 1942; Berger & Pribram 1993). Our brain does not rank probabilities of outcomes, because situations are unique, but rather definite outcomes are manipulated in brain as probability distribution of intervals between excitations, in wave-like way. Overall metric of excitations however is responsible for computations, and when it changes, we would infer $F$ changes to compensate.

Let us introduce notation for conciseness; "<u>where it does not matter what we write, we write nothing at all</u>" (Menger 1952). If distinguishable "slips" $A = \boxed{Abc...z}$, $B = \boxed{aBc...z}$, $C = \boxed{abC...z}$, and so on, which are never found to be combined (they do not overlap), where $\boxed{X} = \boxed{XX}$, $\boxed{XY} = \boxed{YX}$, $\boxed{Xx} = $ o (e.g., Jevons 1870; Yule 1912), are equivalence-classes of physical space $E^*$, namely, different labels mapped to different points of excitation in cerebral cortex, then $I(A,B) = I(B,A)$ is probability that excitation $A$ is not-distinguishable from $B$ *in its ability to change $\alpha$ or $\beta$*, so that, $I(A,A) = 1$, $I(A,B) \cdot I(B,C) < I(A,C)$;

if we wish to perceive "nearness" and "distance," in terms of effect upon $\alpha$ or $\beta$, then $-\log I(A,B) = \Delta(A,B) = \Delta(B,A)$, so that $\Delta(A,A) = 0$, $\infty > \Delta(A,B) > 0$, $\Delta(A,B) + \Delta(B,C) \geq \Delta(A,C)$;

if we wish to construct probabilistic ratios of measurable variable-quantities, e.g., $I((A,B),C)$, $I((A,B),(C,D))$, $I(((...),...),...)$, $\Delta((A,B),C)$, $\Delta((A,B),(C,D))$, $\Delta(((...),...),...)$, and so on, or if our knowledge of relations between elements of metric spaces $S'$ and $S''$ (which are sets of elements between which "nearness" and "distance" is actually defined) we may nevertheless determine whether metric spaces $S'$ and $S''$ are identical or different by measuring frequencies; they are distinguishable if, for any ordered pair $(\alpha, \beta)$ of elements $\alpha \in A$, $\beta \in B$, $\alpha, \beta \in S', S''$, cumulative distribution function $\Delta_{AB}'$ (which generates $S'$) and cumulative distribution function $\Delta_{AB}''$ (which generates $S''$) are distinguishable;

$\Delta(X; A, B)$ is probability that distance from element $A$ to element $B$ is $< X$, so that $\Delta(X; A, A) = 1$, $X > 0$ (distance between any element and any identical or not-distinguishable element, such as itself, is 0, which is less than any given positive distance $X$ whatever);

$\Delta(X; A, B) = 0$, $X \leq 0$, distance between any given element and any different or distinguishable element is neither negative, nor 0 (which is distance between any given element and any identical or not-distinguishable element, which is itself); $\Delta(X; A, B) = \Delta(X; B, A)$, so that distance from $A$ to $B$ and distance from $B$ to $A$ are identical (Menger 1942; Menger 1951a; Menger 1951b).

If a person behaves differently after environment $Xy$ is substituted for environment $xY$, they prefer $Xy$ to $xY$ or they prefer $xY$ to $Xy$, written $[V(Xy) > V(xY)] \cup [V(Xy) < V(xY)]$; if a person behaves identically after environment $Xy$ is substituted for environment $xY$, they do not distinguish environment $Xy$ from environment $xY$, written $V(Xy) = V(xY)$. If space of physical environments is $E^*$, and space of perceived environments is $E$, then: composition $E = ME^*$ is restriction of $E^*$ to $E$ by injection $E^*$ into neuronal model $M$; composition $Б = ME$ is restriction of $E$ to space of behaviors by injection $E$, again, into $M$ (Pribram 1958; Sokolov 1960a; Sokolov 1960b). Supposing $\Delta(X; A, B)$ describes distances in geometry of $2^N$ such elements all stochastically scattered according to another probability-density function, neuronal model $M$ is, at one level, physical dynamics realizing edge-function $G_R$, according to which, elements whose "distance" $X$ from each other is $X < R$ are connected and excite each other (Penrose 2003). Composition $G_R \Delta(X; A, B)$ must thus also tell us how to specify step-function $F$ according to internal observation route, which describes channels and connections realizing $M$.

## §3

Sums are defined for generalized functions. If proper time $\tau$ is fixed, then demand for any group of people is pre-sheaf of elements in space $E$ corresponding to sequence $\sum_{Ч,\tau} [\lim_{\sigma \to 0+} F^+_{Ч,\tau} \varphi^{-1}_{Ч,\tau} \alpha_{Ч,\tau} - F^-_{Ч,\tau} \phi^{-1}_{Ч,\tau} \beta_{Ч,\tau}]$ in union with every restriction homomorphism

$$[\lim_{\sigma \to 0+} F^+_{Ч,\tau} \rho' \varphi^{-1}_{Ч,\tau} \alpha_{Ч,\tau} - F^-_{Ч,\tau} \rho' \phi^{-1}_{Ч,\tau} \beta_{Ч,\tau}] \xrightarrow{R^{\rho'}_{\rho''}} [\lim_{\sigma \to 0+} F^+_{Ч,\tau} \rho'' \varphi^{-1}_{Ч,\tau} \alpha_{Ч,\tau} - F^-_{Ч,\tau} \rho'' \phi^{-1}_{Ч,\tau} \beta_{Ч,\tau}]$$

which preserves hierarchical order of values, and where proper time $\tau$ in effect determines resolution and must be selected to be sufficiently small, and $\rho''$ is contained in $\rho'$. Individuals are obtained, then, by taking cross-sections. Is demand a sheaf, which would allow further analysis, by guaranteeing invariance under wide range of morphisms? That depends on whether sequence of stimuli and responses is "exact" sequence.

Degrees of freedom are different ways which ambiguously maximum satisfaction, but are not locally determined in society: behavior of any one person affects behavior of any other person, changing behavior required for both to obtain maximum satisfaction (Rashevsky 1958). If every possible sequence $0 \to E^* \to Ч_Ц \to ... \to Ч_Щ \to 0$ of injections $\to$ between different people $Ч_1, ..., Ч_Ц, ..., Ч_Щ \in П$, $[1 = Ц = Щ] \cup [1 < Ц < Щ] \in \mathbb{R}$, is "exact," because $[\text{Im } Ч_{Ц-\upsilon} = \text{Ker } Ч_Ц] \cap [\text{Im } Ч_Ц = \text{Ker } Ч_{Ц+\upsilon}]$, then behavior of $Ч_Ц \cap ... \cap Ч_Щ$, which is group of different people, is "smooth," and degrees of freedom are not reduced.

If exactness of such sequences exists, then $H_{\{Э\},P}(E, Э)$ is $P$-th sheaf co-homology group of space $E$ with coefficients in sheaf $Э$, which refers to such sequences. That is, "smooth" behavior of many people means they are "society" of people. In the inductive limit, we may consider, demand and supply generated by many people as hyper-function defined on real-manifold $\mathcal{M}$ comprising of sum of $F_Ч$ where $Ч = \{..., Ч' \in П, Ч'' \in П, Ч''' \in П, ...\}$ (e.g., Sato 1960).

In any such "exact" sequence, output (response) of any one person is input (stimulus) of any other person for which other person also has response. If other person has no response for response of previous person to some input, then society breaks-down at that point. There are situations where degrees of freedom are lost, e.g., violence, because no response to violence will create stimulus to next person for which next person has solutions which do not diminish their own degrees of freedom, in this sense. Division of labour is more productive than isolated work, but any loss of exactness in any such sequence would, among other things, diminish division of labour and extent of market. Exactness is measure of peaceful cooperation of different people, based on division of labour, whereby different preferences requiring work of many different individuals to satisfy are simultaneously satisfied. If one person has no response to stimulus which is itself response of another person, then emotional interruption of behavior, as defined earlier, or violence, will begin at that point (Hebb 1949).

Social conventions, such as division of labour, which are invariant to particular values individuals hold, satisfy exactness of any such sequences, because whatever individuals decide to do, such social conventions are more productive than alternatives, such as isolated labour. Supply arises in this way. However, nothing in $M \in Ч$ makes [Im $Ч_{Ц-υ}$ = Ker $Ч_Ц$] ∩ [Im $Ч_Ц$ = Ker $Ч_{Ц+υ}$] certain to hold, and most social conventions do not result in exactness of sequences. Human society, if compared to animal group in general, is exceptional when it has social conventions which result in exact sequences.

In animal group, individuals quickly learn (by attacking one another) to avoid and not retaliate against some individuals and to not avoid and retaliate against other individuals, resulting in domination directed downward along ranking and submission directed upward along ranking. In animal group want satisfaction of higher ranking individuals is obtained without voluntary cooperation and at expense of preference satisfaction of lower ranking individuals, and only preferences which may be satisfied by working alone or by attacking, chasing away, or stealing property from lower ranking individuals are satisfied at all. In monkey group, for instance, domination and submission ranking of every one individual relative every other individual is quickly established in any environment through aggression of one against other directed downward from higher ranking individuals at lower ranking animals, and lower-ranking individuals avoid or neither retaliate nor resist against violent behavior of higher-ranking individuals; out-group individuals either were totally dominated by all in-group individuals and obtained lowest ranking, by learning to avoid or not-retaliate against all in-group individuals, or out-group individuals totally dominated all in-group individuals and obtained highest ranking, by learning to attack all in-group individuals, never anything in between; removal of amygdala usually disrupted formation of domination habits, and caused constant fighting, until such habit were re-learned (Brody & Rosvold 1952; Pribram 1954). This is a very distant picture from what human society looks like, in modern times at least.

If $A = B$, and you substitute $A$ for $B$, you do not get anything more than you started with, and because everything may be expressed as dynamical equations through equality, there is conservation of energy. Production is only creation of value by transformation of matter from one form to other, more useful form. Production is not creation of matter. If we define behavior $Б^*$ and time $\tau$ as equivalent to some boundary transitions, their role as inputs to production and as stimuli may be investigated.

If $\mathcal{U}' \cdot \mathcal{U}'' = 0$, $Q \in \mathbb{R}$, economic goods are inputs and outputs, and value of factors of production derives from value of their product, which is labeled as being of "lower order" $O^J$ or "nearer to consumption" (Menger 1871), then neither domain nor range of production process

$$\ldots + Q_{O',\xi'} \cdot F\xi' \cdot \mathcal{U}' + Q_{O'',\xi''} \cdot F\xi'' \cdot \mathcal{U}'' + \ldots = \mathcal{M}(\ldots + Q_{O'+\upsilon,\xi'} \cdot F\xi' \cdot \mathcal{U}' + Q_{O''+\upsilon,\xi''} \cdot F\xi'' \cdot \mathcal{U}'' + \ldots)$$

are "created." This function is computed by previous analysis, and coefficients are determined by technology $\mathcal{M}$. They are elements of complex space, mapped in different ways onto itself. Topological space is not itself changed by different mappings from one part of it to any other part of it. In fact, only new arrangements (e.g., of neural network), or ideas, can be created. There are infinitely many such alternatives and these are functions $\mathcal{M} \in M$, physiologically.

Technology is recipe, but it is also capital good, because time must be spent acquiring it, and skills must be learned over to time to make use of it. Part of present income is thus spent to produce future consumption. Part of present income must therefore be "saved," or allocated in this way. There is always some part of present income which is consumed, else people would never consume, which contradicts definition of having preferences. Parts of "influence structure" of this function which comprises of first partial derivatives gives marginal contribution of each factor of production to each product, and thus determines its value. We may abbreviate this by writing $\mathcal{M}(Б^*, I^*)$, when we decode stimuli labels arising from behavior of people.

We can "discover" new functions or mapping between different parts of topological space. We can physically create new technology to produce economics goods, but when we produce economics goods, we do not physically create anything new. There are many ways to produce anything, and from this, we derive concept of "substitution"; and many technologies are only useful as sequences, not individually, from which we may derive concept of "complementary" ideas (Jewkes et al 1958). That is, $[\mathcal{M}'+\mathcal{M}''+...](Б^*, I^*)$ is defined.

Social conventions so fixed, supply is $\sum_{ч,\tau} \mathcal{M}_{ч,\tau}(Б^*_{ч,\tau}, I^*_{ч,\tau}) \in E^*$, with coefficients in $\mathbb{R}$, and again, we may investigate this with differential operators. This visually is kind of phase-space, where different results of different behaviors are different areas and have multiple ways of being reached, and reflects distribution of different economics goods available in society.

Most work which is not trivial, then, goes into determining internal statistical indicators $\alpha$ and $\beta$ for different individuals. Much of interpretation depends on how we select coverings of $E^*$, which are equivalence-classes $\{\xi_J\}[N=\{-\infty,...,K'\}]$, $\{\xi_J\}[N=\{K'',...,K'''\}]$, …, $\{\xi_J\}[N=\{K'''',...,\infty\}]$, and so on, into which stimuli are grouped. Results of behaviors corresponding to such different equivalence classes are different "industries." Results of analysis, again, need only to be computed for any one restriction to be valid for all of them.

### §4
### Conclusion.

Mathematical basis which applies to choice-rich situations of infinitely many different stimuli, where pleasure is state of mind and so we do not arbitrarily classify preferences as selfish or not selfish, but merely as changes of state of mind as responses to different hierarchically-ordered stimuli, and where all existing true propositions of economics follow from mathematical basis itself without additional extra-mathematical assumptions, is presented.

Leonid A. Shapiro    December 31, 2011    arxiv.org/abs/1112.2630    page 22

Models of decision-making derived from such new mathematical basis are not only consistent with results of all experiments in physiology so far, but also quantitatively relate experimentally obtained physiological observables with predictions of economic models so constructed: variable-quantities in such models precisely correspond to observables which determine statistical properties of central nervous system. We wish to model decision-making in our world as it exists, to relax assumptions which prevent our model of decision-making from applying to actual decision-making in our world; we cannot assume, therefore, for instance, that order of consumption of different things is fixed before decisions to consume bundles of very many different quantities of things are made or contents of bundles are determined. We must be able to fill out our model by measuring physiological observables; if we do not incorporate physiological observables into our model, many routes of investigation of human decision-making are not available, and we cannot smoothly pass from model of how brain is working to model of decision-making. If physiological observables fit fundamentally into our model of decision-making, new experimental, mathematical, statistical methods of investigation are found to become available, new graphical methods for quick manipulation of large amounts of observational data become possible, and existing knowledge is re-expressed in robust way, increasing number of aspects of decision making open to analysis.